\documentclass[a4paper,11pt]{article}
\usepackage{pos}

\title{The Tracking performance for the IDEA drift chamber}

\author*[a]{W. Elmetenawee}
\author[b]{G. Chiarello}
\author[c]{A. Corvaglia}
\author[c,d]{F. Cuna}
\author[a,e]{N. De Filippis}
\author[c,d]{E. Gorini}
\author[c]{F. Grancagnolo}
\author[a]{M. Maggi}
\author[c]{A. Miccoli}
\author[c,d]{M. Panareo}
\author[c]{M. Primavera}
\author[a]{G. F. Tassielli}
\author[c,d]{A. Ventura}

\affiliation[a]{Istituto Nazionale di Fisica Nucleare Sezione di Bari, Via Giovanni Amendola, 173, 70126 Bari, Italy}
\affiliation[b]{Istituto Nazionale di Fisica Nucleare Sezione di Pisa, Largo B. Pontecorvo 3, 56127, Pisa, Italy}
\affiliation[c]{Istituto Nazionale di Fisica Nucleare Sezione di Lecce, Via Arnesano, 73100 Lecce, Italy}
\affiliation[d]{Dipartimento di Matematica e Fisica "Ennio De Giorgi" - Universitá del Salento, Via Arnesano, 73100 Lecce, Italy}
\affiliation[e]{Politecnico di Bari, Via E. Orabona, 4, 70125, Bari, Italy}

\emailAdd{walaa.elmetenawee@ba.infn.it}

\abstract{The IDEA detector concept for a future e$^{+}$e$^{-}$ collider adopts an ultra-low mass drift chamber as a central tracking system. The He-based ultra-low mass drift chamber is designed to provide efficient tracking, a high-precision momentum measurement, and excellent particle identification by exploiting the cluster counting technique. This paper describes the expected tracking performance, obtained with full and fast simulation, for track reconstruction on detailed simulated physics events. Moreover, the details of the construction parameters of the drift chamber, including the inspection of new material for the wires, new techniques for soldering the wires, the development of an improved schema for the drift cell, and the choice of a gas mixture, will be described.}

\FullConference{%
  41st International Conference on High Energy physics - ICHEP2022\\
  6-13 July, 2022\\
  Bologna, Italy
}

\begin{document}
\maketitle

\section{The tracking requirements for FCC-ee}
The FCC-ee circular electron-positron collider is designed to study the Z, W, and Higgs bosons and the top quark with high precision by producing samples of $5 \times 10^{12}$ Z bosons, $10^{8}$ Wpairs, $10^{6}$ Higgs bosons and $10^{6}$ top quark pairs. The high performance is expected to be achieved by utilising the experience gained with LEP at high energies and the high luminosity features developed on the b-factories. An overview of the proposed detector layout and performance requirements can be found in the FCC-ee Conceptual Design Report (CDR)~\cite{a}. The tracking system, which makes up the innermost part of the FCC-ee detector, must be capable of delivering outstanding performance across the full acceptance, down to approximately 120 mrad, and full momentum range, typically with efficiencies of about 98\% down to 300 MeV/c or better for muons down to 100 MeV/c transverse momentum. Furthermore, state-of-the-art momentum and angular resolution for charged particles are required, with a target of $\sigma (1/p_{T}) \leq 3\times 10^{-5} (\text{GeV}/c)^{-1}$ and  $\sigma(\Theta,\phi)\approx 0.1$ mrad for 45 GeV muons, driven by the requirements for precise measurements of the Higgs boson.

The magnetic field will be limited to approximately 2 T, in order to contain the vertical emittance at the Z pole, and a tracking volume up to a relatively large radius is needed to recover momentum resolution. It must be engineered in a way that results in minimum material in front of the external detectors and a stable structure capable of providing measurements with a low enough systematic error to match the enormous statistics expected, particularly for the Z pole running. Particle identification capability would also be a valuable additional ability. For the innermost vertex layers, these goals imply a target hit resolution of about 3 $\mu \text{m}$ along with a material budget of the innermost layer of the vertex detector of 0.2\% of the radiation length or better, with a target of less than 1\% for the entire vertex detector. 

\section{The IDEA detector concept and its tracking system}

\begin{figure}[ht]
\centering
\includegraphics[width=9.5cm]{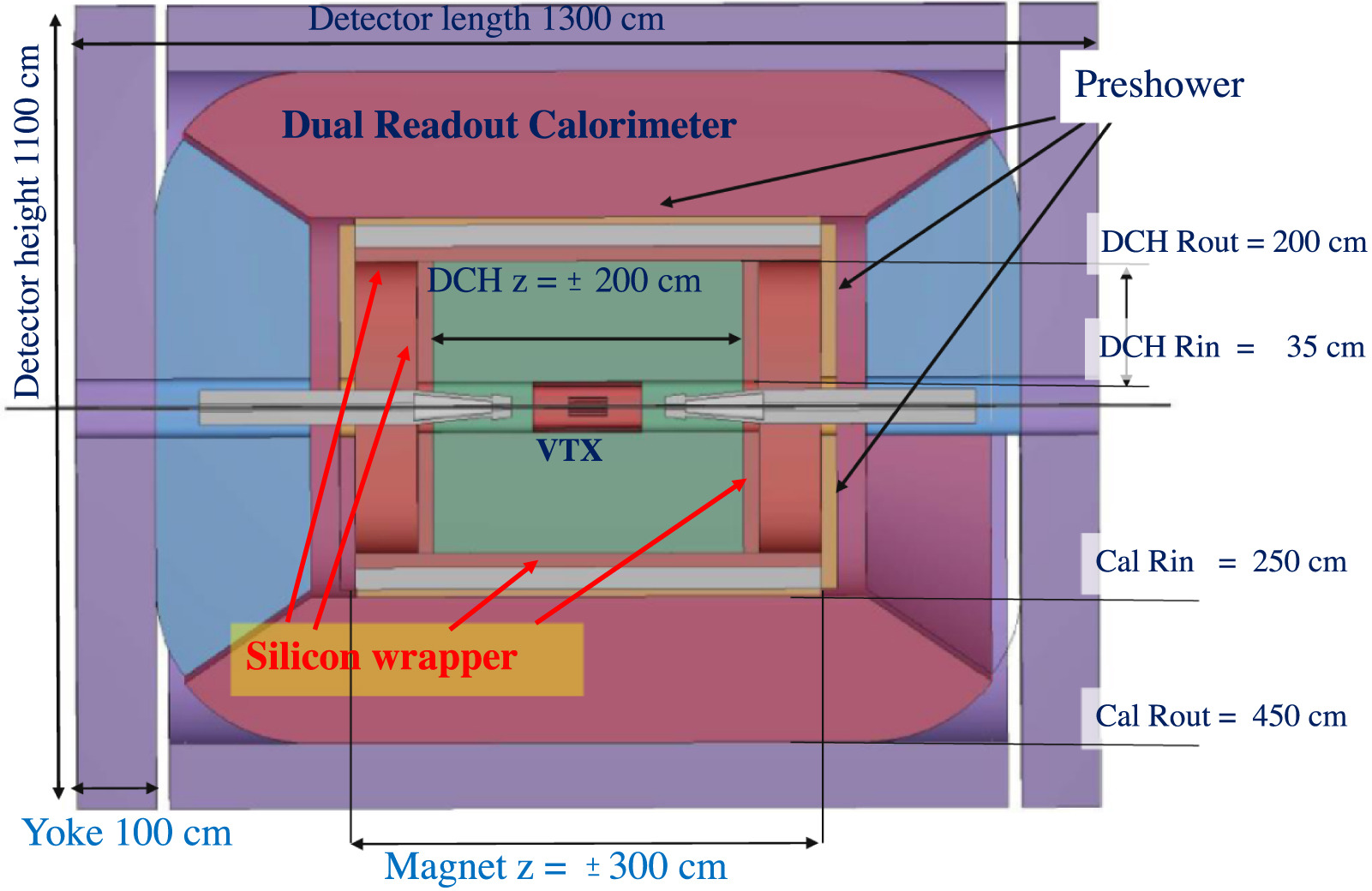}
\qquad
\caption{Schematic layout of the IDEA detector.}
\label{fig:IDEA}
\end{figure}

The IDEA detector (Innovative Detector for Electron-positron Accelerator)~\cite{b} is a general-purpose detector designed for experiments at the FCC-ee collider. The current design of the IDEA detector concept,  outlined in Figure~\ref{fig:IDEA}, features a silicon pixel detector, a large-volume drift chamber surrounded by a layer of silicon micro-strip detectors, a superconducting solenoid coil, a preshower detector, a dual-readout calorimeter and a muon spectrometer within the magnet return yoke. The tracking system is based on all the detectors inside the calorimeter. A key element in the structure of IDEA is an ultra-thin ($\sim$30 cm thick) and low mass ($\sim$0.7 X$_{0}$ and $\sim$0.16 $\lambda_{int}$) solenoidal coil with a magnetic field of 2 T located between the tracking and the calorimeter volume. The low magnetic field minimises the impact on emittance growth and allows for manageable fields in the compensating solenoids. Moreover, positioning the coil inside the calorimeter volume reduces the stored energy by a factor of four, and the cost can be halved.

The vertex detector, the innermost detector surrounding the 1.5 cm beam pipe, is a silicon pixel detector based on monolithic active pixel sensors designed for precisely determining the impact parameter of charged particle tracks. Technology relying on fully depleted high-resistivity substrates is being considered, together with implementing on-pixel sparsification and data-driven, time-stamped readout schemes. The target performance would be a resolution of about a few $\mu$m, thickness in the range of 0.15–0.30\% X$_{0}$ per layer and power dissipation not exceeding 20 mW/cm$^{2}$ in order to avoid the need for active cooling.

The central tracker of the IDEA detector is based on a very light central drift chamber (DCH), which should achieve lower mass than the equivalent silicon-based tracking and provide good tracking and high-precision momentum measurement. A novel feature of this detector is that it is instrumented with readout electronics implementing the cluster counting/timing techniques, allowing for excellent particle identification over most of the momentum range of interest. The DCH will be discussed in more detail in the next section.

The preshower detector is based on the micro-Resistive WELL ($\mu$-RWELL) technology~\cite{d}: a compact Micro-Pattern Gaseous Detector (MPGD) with a single, intrinsically spark-protected amplification stage. In the barrel region, the magnet coil works as an absorber of about 1 X$_{0}$ and is followed by a layer of MPGD $\mu$-RWELL chambers; the second layer of chambers follows after another 1 X$_{0}$ of lead. A similar construction exists in the forward region, with both absorber layers made from lead. Besides increasing the overall tracking resolution, the $\mu$-RWELL chamber layers provide a precise acceptance determination for charged particles.

\section{The IDEA central drift chamber}

\begin{figure}[ht]
\includegraphics[width=8cm]{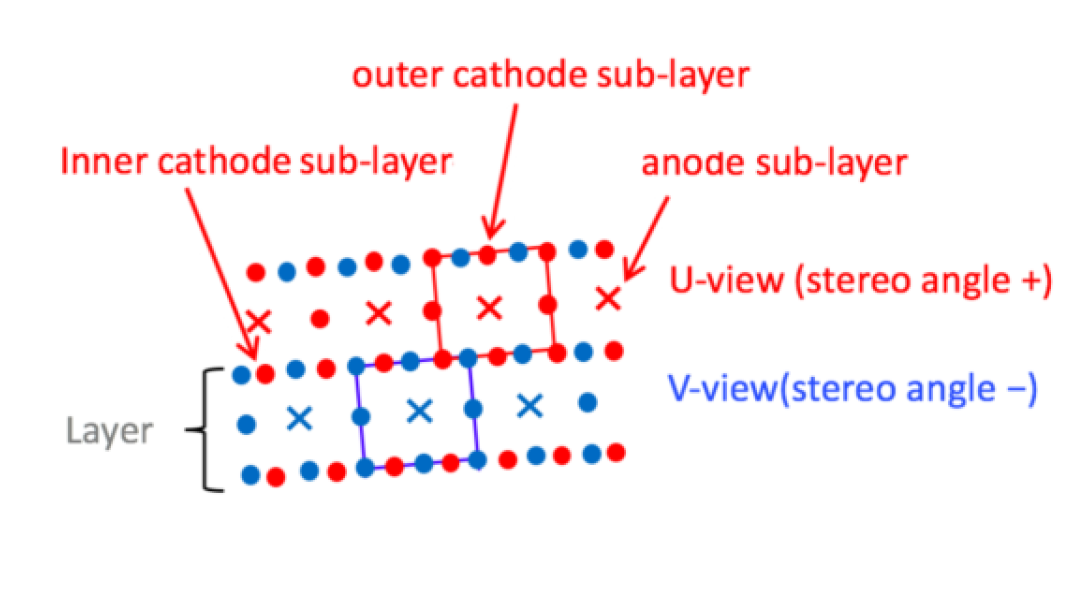}
\qquad
\includegraphics[width=6cm]{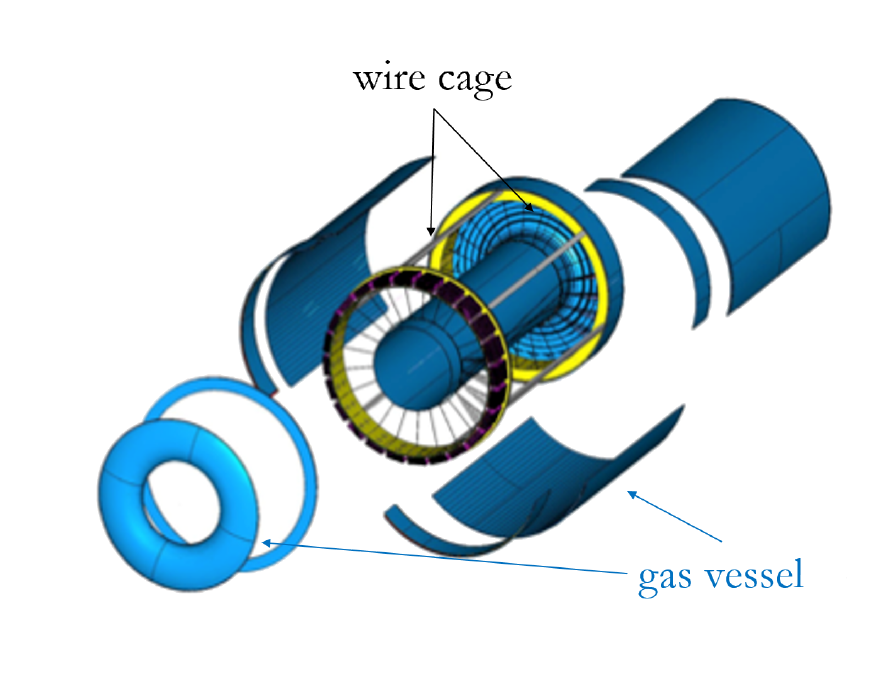}
\caption{A sketch of the drift cells within two alternating sign stereo layers (left). Schematic layout of the CDCH mechanical structure (right).}
\label{fig:DCH}
\end{figure}

The IDEA Central Drift chamber (CDCH) is a unique-volume, high granularity, fully stereo, low-mass cylindrical chamber, co-axial with the 2 T solenoid field, operating with a helium-based gas mixture, see Figure~\ref{fig:DCH} (right). The CDCH is 400 cm long, with an inner (outer) radius of 35 (200) cm. The superior feature of this detector is its high transparency in terms of radiation lengths: the total amount of material in the radial direction is about 1.6\% X$_{0}$, reaching about 5\% X$_{0}$ in most of the forward regions. The wires are positioned in a way which enmeshes the positive and negative stereo angle orientations, giving a high ratio of the field to sense wires equal to 5:1 to ensure the proper electrostatic configuration. The sensitive volume comprises 56448 squared drift cells with an edge between 12 and 14.5 mm. Each square cell comprises one anode and two cathode sub-layers, as sketched in Figure~\ref{fig:DCH} (left); The anodes are 20 $\mu$m in diameter tungsten wires, while the cathodes are 40 and 50 $\mu$m light aluminium alloy wires. Overall, the CDCH is made with 56,448 sense wires, 285,504 cathode wires and 2,016 guard wires to equalize the gain of the innermost and outermost layers, requiring non-standard wiring and assembly procedures, inspired by the MEGII drift chamber concept~\cite{e}. 

The gas mixture chosen to operate the CDCH is 90\% He and 10\% $\text{iC}_{4}\text{H}_{10}$, chosen for the low radiation length ($\sim$1400 m), a fast enough average drift velocity ($\sim$2 cm/$\mu$s) corresponding to a maximum drift time less than 400 ns, and a good spatial resolution (110 $\mu$m, [6]). The number of ionization clusters generated by a minimum ionizing particle in this gas mixture is about 12.5 $\text{cm}^{-1}$, allowing for exploiting the cluster counting/timing techniques for particle identification with unprecedented resolutions~\cite{aa}. Typically it is expected to be a factor of two better than the traditional dE/dx technique and to improve the spatial resolution obtainable with the conventional measurement of the fastest drifting electron to well below 100 $\mu$m for short drift cells.~\cite{f}.

\section{The Expected Tracking performance}

The momentum resolution of the IDEA tracking system was evaluated with a Geant4 simulation and an analytical approach. Assuming a single-cell resolution of 100 $\mu$m for the CDCH, as expected~\cite{f}, and conservative spatial resolution (pitch/$\sqrt{12}$) for Si detectors, the IDEA tracking system fulfils the expected performance. Moreover, the resolution at high momentum, about 100 GeV/c, can be improved from $4\times 10^{-5}$ to $2.9\times 10^{-5}$  using a less conservative spatial resolution for the Si detectors. 

\begin{figure}[ht]
\includegraphics[width=8.5cm]{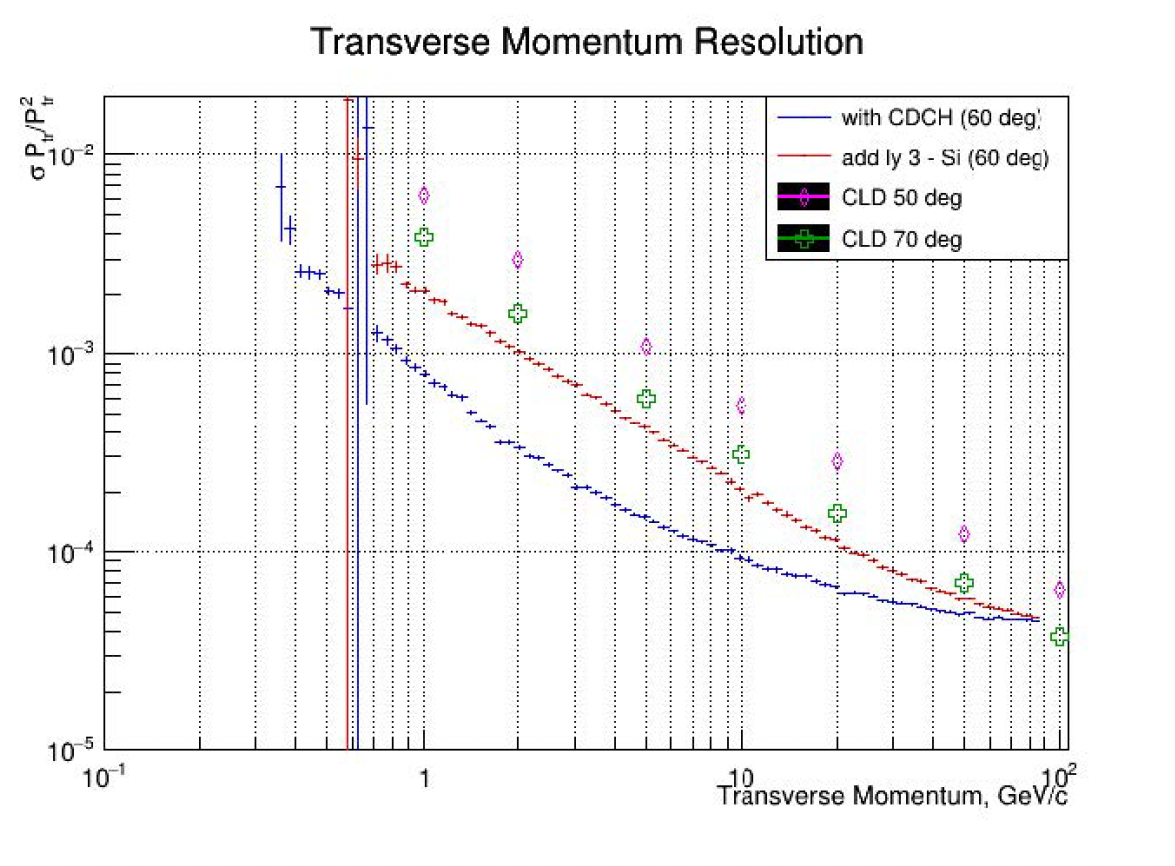}
\quad
\includegraphics[width=7.1cm]{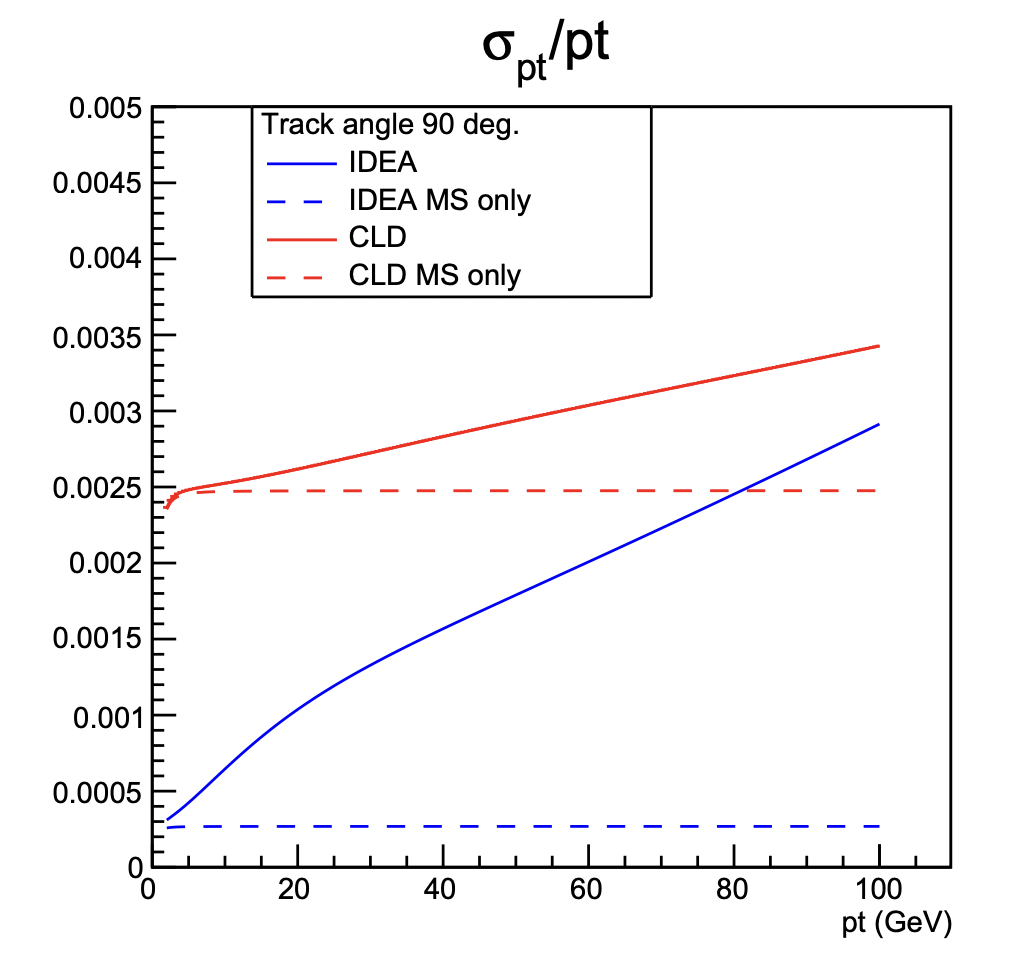}
\caption{Transverse momentum resolutions of the IDEA tracking system and a full Si-based tracking system similar to the one of the CLD detector as obtained with Geant4 simulation (left) and an analytic calculation of the full covariance matrix that takes into account spatial resolution and material effects (right).}
\label{fig:perform1}
\end{figure}

The lightness of the drift chamber results in a gain in the momentum resolution of the IDEA tracking system, quantified in about a factor of 3  with respect to a full Si tracker system up to 50 GeV/c~\cite{g} as shown in Figure~\ref{fig:perform1} (left). The dashed lines in Figure~\ref{fig:perform1} (right) exemplify the multiple scattering contribution~\cite{h}. Due to the minimization of the multiple scattering, the IDEA tracking system performs better than an alternative tracking system based only on Si detectors over almost the entire momentum range of interest, as shown in Figure~\ref{fig:perform1}.
\bigskip

\begin{figure}[ht]
\includegraphics[width=8.2cm]{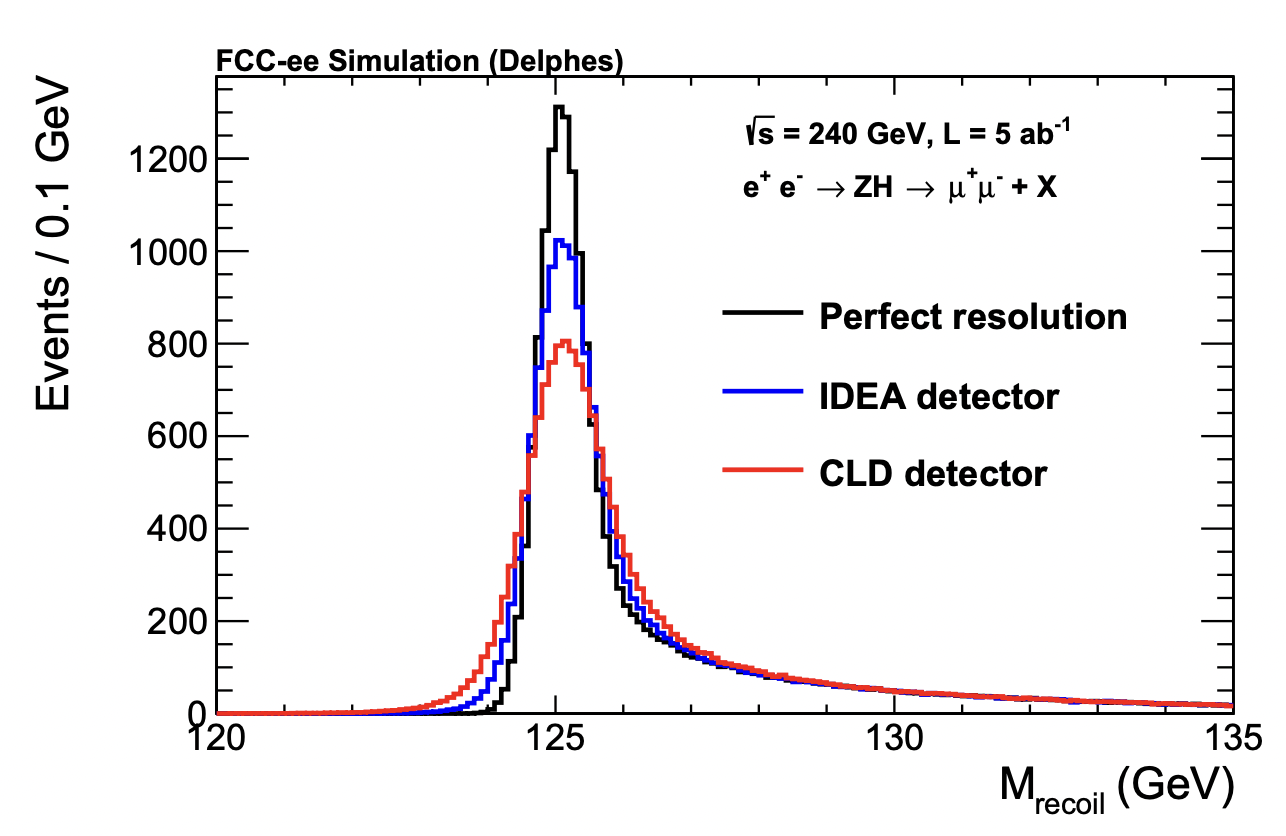}
\quad
\includegraphics[width=7.8cm]{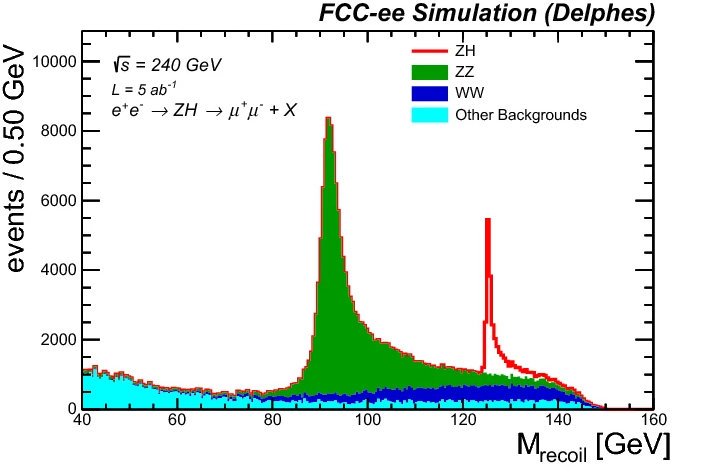}
\caption{Distribution of the Higgs recoil mass in ZH events where the Z decays to $\mu^{+}\mu^{-}$ in the region around the Higgs mass (left), assuming: an ideal momentum resolution (black), such that the resolution on the recoil mass is determined by the beam energy spread; the momentum resolution of the IDEA detector (blue); that of the CLD detector (red). The inclusive recoil mass distribution between 40 and 160 GeV (right), displaying the Z peak from the ZZ background (green) and the H peak from the ZH signal (red).}
\label{fig:perform2}
\end{figure}

A key measurement of the $e^{+} e^{-}$ Higgs factories is the total cross-section of ZH production; This is done by measuring the Z momentum and analyzing the recoil mass while using the collision energy as a constraint. In the case of circular colliders, the beam energy spread (BES) is minimal, $\Delta E/E \sim 10^{-3}$; thus, spoiling it with poor detector resolution should be avoided. In Figure~\ref{fig:perform2} (left), the Higgs recoil mass in ZH events, where the Z decays into muons, expected with perfect knowledge of the Z momentum, is shown and compared to the expectations with the IDEA and CLD tracking systems. The goal is that the reconstruction of the recoil mass is limited by BES rather than by the detector resolution. The very light tracker from IDEA, with a resolution of 0.15\% for central, 50 GeV muons, is close to reaching this goal~\cite{i}. The (heavier) full silicon tracker of CLD performs a bit worse because the resolution is dominated by multiple scattering in the momentum range of interest.

The determination of the Higgs mass, for which a precision of a few MeV would be needed in view of a possible run at the Higgs resonance, will clearly benefit from the better momentum resolution offered by a light, gaseous tracker. This is illustrated in Figure~\ref{fig:perform2} (right), which shows the recoil mass distribution, obtained with a DELPHES simulation of the IDEA detector concept, in particular its drift chamber, for an integrated luminosity of 5 ab$^{-1}$ simulated at $\sqrt s$ = 240 GeV and with a nominal Higgs boson mass of $m_{H}$=125 GeV~\cite{j}. The large signal-to-background ratio on the one hand and the excellent muon momentum resolution of the drift chamber on the other offer the possibility to determine the inclusive ZH cross section and the Higgs boson mass with a statistical precision of $\sim$1\% and $\sim$6 MeV, respectively.

\end{document}